\documentclass[twocolumn,showpacs,preprintnumbers, superscriptaddress,amsmath,amssymb]{revtex4}

\usepackage{graphicx}
\usepackage{dcolumn}
\usepackage{bm}
\usepackage{color}

\usepackage{hyperref}

\begin{document}

\preprint{APS/123-QED}

\title{Active Motion of Janus Particle by Self-thermophoresis in Defocused Laser Beam}

\author{Hong-Ren Jiang}
\affiliation{%
Department of Physics, The University of Tokyo, Hongo 7-3-1, Tokyo 113-0033, Japan
}
\author{Natsuhiko Yoshinaga}
\affiliation{
Fukui Institute for Fundamental Chemistry, Kyoto University,
Kyoto 606-8103, Japan
}
\author{Masaki Sano}%
\affiliation{%
Department of Physics, The University of Tokyo, Hongo 7-3-1, Tokyo 113-0033, Japan
}%

\date{\today}

\begin{abstract}
We study self-propulsion of a half-metal coated colloidal particle under laser irradiation. 
The motion is caused by self-thermophoresis: i.e. 
absorption of laser at the metal-coated side of the particle creates local temperature gradient which in turn
drives the particle by thermophoresis.
 To clarify the mechanism, temperature distribution and a thermal slip flow field around a micro-scale Janus particle 
are measured for the first time.
With measured temperature drop across the particle, the speed of self-propulsion 
is corroborated with the prediction based on accessible parameters.
As an application for driving micro-machine, a micro-rotor
 is demonstrated.
\end{abstract}

\maketitle

Self-propulsion arouses interests among scientists because of its
interesting dynamics, associated theoretical challenges to
nonequilibrium transport phenomena, and possible applications, including
micro-swimmer, nano-machine and drug delivery \cite{1,2,3,4}. 
Biological self-propelled molecules or biomotors \cite{7}, such as 
ATPase and myosin, convert chemical energy to mechanical motion through chemo-mechanical coupling.  Such thermodynamic coupling 
also causes self-propulsion in physical systems as exemplified by 
nano-rods motion in hydrogen peroxide solutions \cite{8}. 
In contrast to self-propulsion, phoretic motions, such as
electrophoresis, dielectrophoresis and thermophoresis, generally
described by the linear response theory, are the directional motions of
material in given external fields. 
To make a bridge between a phoresis and a self-phoretic motion \cite{9} would give
more insights into the mechanisms for self-propulsion. In this paper, 
 we produced Janus particles (the particle containing two different surfaces
on its two sides \cite{Janus}), by evaporating gold coat on the hemisphere of silica 
or polystyrene spheres, and analyzed self-propulsive motion under the
laser irradiation. The Janus particle moves along a local temperature
gradient which is generated by absorption of laser at the metal-coated
hemisphere. 
The local driving gradient and flow near a micro-scale Janus particle are experimentally observed for the first time.

\begin{figure}
\includegraphics[width=8.6 cm]{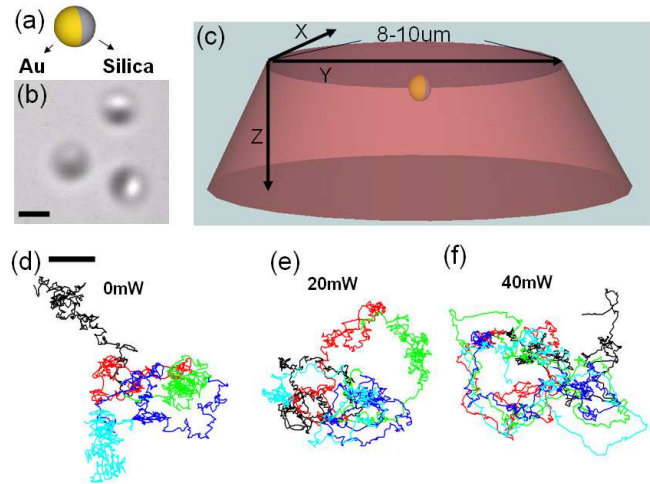}
\caption{\label{fig:epsart1} (color online) (a)(b) Janus particle
 and its bright field image under microscope. The dark sides of the
 particles shown in (b) are the Au coatings (scale bar:1 $\mu$m). 
 (c) Schematic drawing of a
 Janus particle placed in a thin chamber (thickness: 20 $\mu$m). The
 laser, red (gray) cone, is fed in the Z-direction from the bottom and
 the particles are visualized in the XY-plane. 
(d) to (f): The trajectories of a Janus particle in the XY-plane within 10 
seconds. Each color corresponds to an interval of 2 seconds. 
(d) Without laser irradiation. (e) 20 mW, and (f) 40 mW. Bar: 1 $\mu$m.}
\end{figure}

Thermophoresis of colloids in solution is intensively discussed recently
\cite{10,11} motivated by new experimental observations. 
To utilize thermophoresis as a mechanism for self-propulsion of a
particle, it is essential to create a local temperature gradient by the
particle itself. 
To prepare Janus particles, a mono-layer of micro-scale silica or
polystyrene particles is prepared by a drying process. 
 Gold is deposited by thermal evaporation 
to create 25 nm thick coating on the half side of each particle \cite{13}. 
An image of 1 $\mu$m Au-silica Janus particles is shown in Fig. 1(a) and (b). 
A thin chamber containing a solution  and particles is sandwiched by two cover glasses
with a 20 $\mu$m spacer. 
The cover glasses are treated by BSA to prevent particles from sticking
to the surfaces. 
A laser (Nd:YAG, 1064 nm) is fed from the bottom of the chamber through
an oil immersion objective (100X NA 1.35. The laser is defocused to irradiate
a wider area of a diameter about 9 $\mu$m (Fig.1 (c)), where 20\% of maximum intensity is defined as its edge. 
Images of particles are collected through the same objective by a high
speed camera at 300 frames per second. 

1 $\mu$m Au-silica Janus particles are placed in a
chamber at a low concentration to insure only one particle caught within
the view under the microscope. After applying the laser, a particle is
pushed to the upper surface of the chamber and moves around within the
laser irradiated region. 
The two-dimensional trajectories of a particle
are shown in Fig. 1(d)-(f). 
The
particle moves faster under laser irradiation than the normal diffusion
and the direction of motion follows the polarity of the Janus particle
(See movie-1 in Supplementary Information(SI) \cite{12}).

\begin{figure}
\includegraphics[width=8.6 cm]{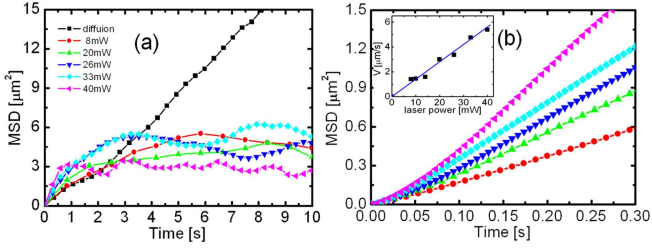}
\caption{\label{fig:epsart2} (color online) (a) MSD of self-propulsion of the Janus particles under different laser power. (b) The MSD in small time scale. Inset: the relation between self-propelled velocity and laser power.}
\end{figure}

We measured the mean square displacement (MSD) of the moving
particle. The MSD for different laser irradiation powers are shown
in Fig. 2(a). The power of the laser is varied from 0 to 40 mW before
the objective. We observed that the irradiation of the laser increases
the MSD several folds for a short time scale ($\leq $1
second) and the confinement of the particle for a longer time scale
($\geq $3 second). The latter can be explained by the
two-dimensional trapping effect due to strong scattering from the metal
part of particle \cite{14}, although the laser is defocused. 
On the other hand, the behavior at the shorter time scale cannot be
attributed by the trapping effect.

Based on these observations, we consider that the  dynamics of a Janus
particle is composed of three kinds of motions, Brownian motion,
self-propelled motion induced by the laser, and optical trapping
confinement. These three kinds of motions can be separated by two
different time scales; the time constant $\tau _k$ of the laser
trapping effect and the rotational diffusion time constant $\tau _r $  of self-propelled motion induced by the laser.

 To discuss the time scale of the trapping effect, we employ the motion
 of a self-propelled particle in a harmonic potential without inertia as a first
 approximation. 
The MSD of the particle can
 be obtained as (See SI),
\begin{eqnarray}
MSD = &
4 D \tau_k ( 1 - e^{-t/\tau_k})
- \frac{2 V^2 \tilde{\tau}}{\tau_k^{-1} + \tau_r^{-1}} e^{-t/\tau_r}
\nonumber \\
& + \frac{2 V^2 \tau_k}{\tau_k^{-1} + \tau_r^{-1}}
\left(
1 + \frac{\tilde{\tau}}{\tau_r} e^{-t/\tau_k}
\right),
\label{Eq_MSD}
\end{eqnarray}
where $\tilde{\tau} = 1/(- \tau_r^{-1} + \tau_k^{-1})$, $D_r = 1/2 \tau_r$ is the rotational
 diffusion constant, $D$ the diffusion constant of the particle, $V$ the self-propelling velocity, $k$ the spring constant,
 $\tau_k = k_BT/D k$ the time constant due to the harmonic potential, $T$ the
 temperature, and $k_B$ is Boltzmann constant. 
When $\tau_k
 \gg \tau_r$, Eq.(\ref{Eq_MSD}) tends to 
$MSD \simeq 
(4 D  + 2  V^2 \tau_r) \tau_k 
(1 - e^{-t/\tau_k}),
$
 thus $\tau _k$ can be obtained
 from MSD and Eq.(\ref{Eq_MSD}) by an approximation. 
The values of $\tau _k$ are
 from 0.4 to 3 s depending on the laser power.

The self-propelled motion along the particle's polarity ${\bf n}(t)$ is characterized by the rotational diffusion of the polarity. 
By measuring polarity of the particle from the images, 
the rotational diffusion time constant $\tau _r$ was obtained by
calculating the correlation function  $\langle {\bf n}(t)\cdot {\bf n}(0) \rangle = \exp(-t/\tau _r)$ 
 (data not shown).
The values of $\tau_r$ are typically fitted  between 0.1 to 0.2 s
depending on the laser power. 
The dependence may come from the jumps of the polarity through the
direction normal to the observing plane, which are more probable under higherlaser power. This leads shorter $\tau_{r}$ obtained by 2D projection of the polarity. In a short time scale, $\ t\ll  \tau _r \ll \tau _k$, the rotational
 diffusion and the trapping effect can be neglected, the motion can be
 treated as Brownian motion with a fixed directed motion. The MSD of Eq.(\ref{Eq_MSD}) in a short time scale becomes, 
\begin{equation}
\ MSD=4Dt+V^2t^2.
\label{Eq_DV}
\end{equation}
The contribution from Brownian motion is $4Dt$ and that of directed
motion is $\ V^2t^2$. Similar behavior has been discussed in
hydrogen peroxide solutions but without potential \cite{15} and in
motility of cell \cite{selmeczi:2005}. 
Eq.(\ref{Eq_DV}) is used to fit the MSD in Fig. 2(b) up to 0.1 s to
obtain a set of different $\it V$ values for different laser powers with
a fixed $D$. 
The data can be well fitted by Eq.(\ref{Eq_DV}) and $\it D$ is obtained as 0.38 $\mu m^2/s$. Comparing with diffusion coefficient of a free particle (0.47 $\mu m^2/s$), smaller value might be caused by the additional friction over the wall. The relation between self-propelling velocity $\it V$ and the laser irradiation power is shown in the inset of Fig. 2(b). The overall linear relation confirms that the self-propulsion of a particle is powered by the laser irradiation linearly. 

\begin{figure}
\includegraphics[width=8.6 cm]{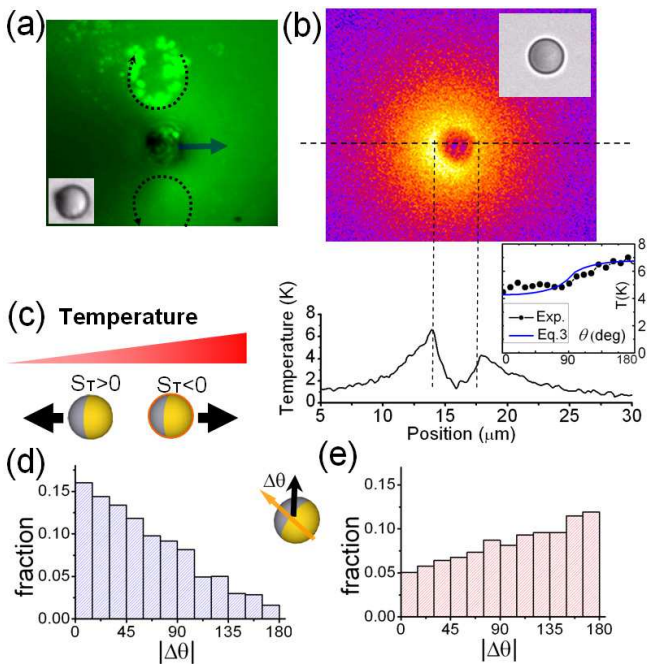}
\caption{\label{fig:epsart3} (color online) (a) Distribution of 40 nm
 fluorescent tracers around a tethered 3 $\mu$m Au-PS Janus particle under laser irradiation. The image is accumulated for 5 s and the
 bright spots are the trace of a larger particle. Dashed line arrows
 indicate the flows induced by temperature gradient and the solid arrow indicates the direction of motion in the non-tethered condition. (b) The temperature distribution for tethered 3 $\mu$m Au-PS Janus particle under laser irradiation. Inset: Surface temperature (solid line is from Eq.(3)). (c) The direction of motion for a Janus particle in water and in a Triton X-100 solution. (d) Directional angle distribution of motion with respect to the polarity in water. Inset, black arrow: direction of motion; orange (grey) arrow: polarity. (e) Same kind of distribution as (d) in the Triton X-100 solution.}
\end{figure}

We propose that the directed motion of Janus particle is induced by the temperature gradient
 generated by itself which we call self-thermophoresis.
The metal side absorbs the laser of the intensity $\it I$ with an absorption efficiency
$\epsilon$. 
The heat generation at the surface of the particles with the radius $R$ is balanced with thermal diffusion thus gives
$q(\theta )=\kappa {\bf e}_n \cdot \nabla T$  where $\kappa$ is the heat
conductivity of water and  ${\bf e}_n$ is the outward normal unit vector
on the surface. 
The temperature distribution on the surface is obtained from
heat transfer equation as
\begin{equation}
\vspace{-.10cm}
 T(R) = 
T_0 + \sum_{n=0}^{\infty}
\frac{q_n R}{(n+1)\kappa_o + n \kappa_i}
P_n(\cos \theta)
\label{Eq_T}
\vspace{-.10cm}
\end{equation}
where $T_0$ is the temperature at infinity and $P_n(\cos \theta)$ is the
Legendre polynominal of order $n$. 
The thermal conductivity inside and outside the particle is $\kappa_i$
and $\kappa_o$, respectively. 
The heat flux $ q(\theta )$ is expanded using the polynomial with
the coefficients $q_n$. 
The temperature is asymmetric around the particle due to half-side
absorption. 
The gap of temperature  across the two sides of Janus particles is
$\Delta T = 3 \epsilon I R/2 (2
\kappa_o + \kappa_i)$.
The propulsive velocity $V$ for a self-thermophoretic
particle is given by a surface integral of effective surface slip
velocity  $v_s = \mu(\theta) \nabla T$ driven by
local temperature gradient \cite{9}.
$\mu (\theta)$ is {\it mobility} characterized by interactions between
the particle and fluids.
On the other hand, exactly the same gradient can be imposed externally in which the velocity is
described by the Soret coefficient, $S_T =D_T/D$ as $V=-D S_T \nabla T$.
Here $D_T$ is the thermodiffusion coefficient.
This suggests that the self-thermophoretic velocity is described not
only by the mobility but also by the Soret coefficient, which has been rather
extensively studied (see SI).
The velocity is
\begin{equation}
\vspace{-.10cm}
\ V=
- D S_T \frac{\Delta T}{3R},
\label{velocity}
\vspace{-.10cm}
\end{equation}
where $S_T$ of Janus particle is interpreted as the average of
the two coefficients of silica beads and those covered uniformly by gold.
In Eq.(\ref{velocity}), $\Delta T$ is proportional to the laser power and
accordingly velocity  $V$ is proportional to it,
which agrees with the MSD measurements. 
To further verify the proposed
self-phoretic mechanism, the flow and temperature around a Janus
particle and the magnitude and sign of Soret coefficient are compared
with observations in the following.

The flow around a spherical colloid in temperature gradient was reported
 recently\cite{16}.  We expect a similar flow
caused by self-thermophoresis might exist around Janus particle. 
We fixed a 3
$\mu$m Au-polystyrene Janus particle on the glass surface and added 40 nm
polystyrene fluorescent particle to the solution as
tracers. 
The distribution of tracers and motion of a large tracer are
shown in Fig. 3(a). The motion of a large tracer clearly shows the
direction of the circulating flow near the heated Janus particle. We
also found that the concentration of tracers is high in the
non-coating side and low in the coating side. It can be
explained by the balance between the drag by the thermal slip flow
induced by Janus particle and the thermophoresis of the tracer
itself. Tracers near the particle surface are driven from colder to
hotter region by the thermal slip and reinjected to the colder
side by the circulation. But the tracers are repelled from the hotter region 
due to positive Soret effect, resulting in
higher density of tracers at the colder side of the Janus particle.

The temperature profile was directly measured for
an Au-polystyrene Janus particle with the diameter of 3 $\mu$m fixed on the cover glass and
subjected to laser heating. 
We used
a temperature sensitive fluorescent dye,
bis( 2- carboxyethy)5, 6 carboxyfluorescein (BCECF) \cite{17}. 
The 
temperature profile is shown in Fig. 3(b), where the laser power is 100
mW  and the diameter of the irradiated region is 25 $\mu$m.  
Here larger irradiated region is adopted to scale up the length scale together with the particle size (from 1$\mu$m to 3 $\mu$m). 
The average laser intensity per unit area $\it I$ is equal to the condition where the laser 
power is 13mW in the 9 $\mu$m-diameter region.
The temperature in the coated side is about 2 K
higher than the non-coated side (Fig. 3(b)). 
The temperature profile
 qualitatively agrees with the analytic result as shown in Fig.3(b).
 For a 1  $\mu$m bead, the temperature profile is hard to obtain
experimentally; the temperature gap $\Delta T$ across the particle is estimated as about 0.7
K based on 3 $\mu$m data. By using the velocity data in Fig.2(b) inset,
we can estimate $S_T$ from Eq.(\ref{velocity}) as 10  ${\rm K}^{-1}$. 
This result is consistent with the Soret coefficient (26 ${\rm K}^{-1}$) of Au-silica Janus particle 
independently measured by single particle tracing \cite{braun:2006}
in an externally applied
temperature gradient using an optically heated chamber\cite{19, 20}.

Recently, it has been shown that the thermophoretic property of
particles can be changed by adding surfactant, which would be absorbed
on the surface of particles \cite{18}.  
The sign of Soret coefficient of Janus
particle is also determined by single particle tracing. 
It shows that the Soret coefficient of Janus particles is positive in
water (moving to a colder region). 
Next, we add Triton X-100 to the solution (0.05 wt.\%). We find that the Soret coefficient of Janus particles become negative in Triton X-100 solution (particles move to warmer region). Fig. 3(c) is the schematic figure of the directions of thermophoretic motion of a Janus particle in a temperature gradient.  

We measured the angle $\Delta \theta$ between the
direction of the motion and the polarity of Janus particles
in water and Triton X-100 solution as shown in Fig.3(d) and (e). 
The motion of a Janus particle is 
directed from metal side to silica side in pure water, implying toward colder side in the local temperature 
gradient created by local heating. 
Thus the direction of self-propelled motion agrees with that of thermophoresis 
in water (see Fig.3(c)(d)). 
When the Soret coefficient becomes negative in Triton
X-100 solution, 
Janus particles moves toward the
metal-coated (warmer) side \cite{22} which confirms that the laser induced
self-thermophoresis  motion does depend on Soret coefficient. Such reversal property of self-propelled motion in consistent with the reversal of phoretic motion further confirms the proposed mechanism. 

\begin{figure}
\includegraphics[width=8.6 cm]{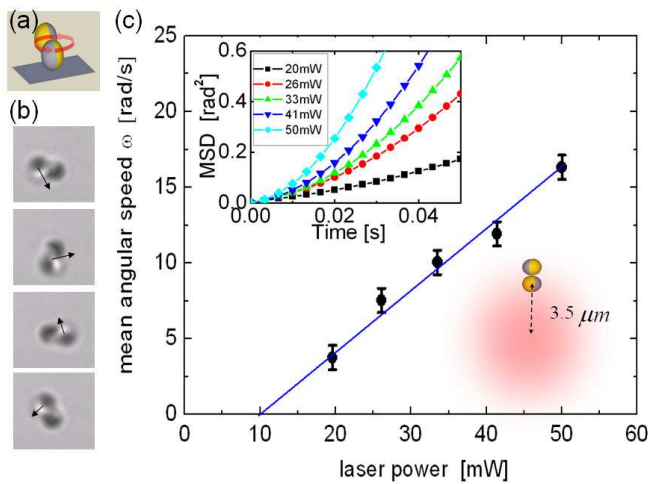}
\caption{\label{fig:epsart4} (color online) (a) Schematic figure and its bright field
 image (b) of a micro-rotor during rotation. 
The time interval in (b) is 0.2 s for successive images. 
The arrow indicates the direction of rotation. 
(c) Rotational speed versus laser power. Inset: the MSD (left) and
 schematic top view of the micro-rotor and laser irradiation (right).}
\end{figure}

Finally we demonstrate that this method can be used to drive a
micro-machine. 
We carefully selected twin Janus particles which consist of
two 1 $\mu$m Au-silica Janus particles with the one being tethered to
the surface 
(Fig. 4(a)). 
We apply a focused laser at 3.5 $\mu$m distance apart
from the tethered particle (the inset of Fig.4(c)). 
As we increased the laser power, we observed that the twin
particles start to rotate. The rotational direction depends on the
polarity of the "vane" Janus particle. The particle moves in the
direction indicated by the arrow. (also see movie-2 in SI). Stable rotations of particles are observed  in the range of
laser power from 20 mW to 50 mW.
Above a certain threshold in the laser power, the rotation speed increases linearly with the laser
power (Fig. 4(c)). 
The threshold for rotation may be due to an interaction potential hill existing
between the bead and the glass surface. The linear
velocity of the vane particle of motor can be calculated from the
rotational velocity. The velocity varies linearly between 2 to 10
$\mu$m/s as a function of the laser power.  
The velocity/power relation evaluated from the slope of Fig. 4(b) is 0.26
$\mu {\rm m}/(s\cdot{\rm mW})$. The rotational motion has a stochastic nature at a low laser power. 
The MSDs of rotor for a short time scale for different laser power are shown in the inset of Fig.4(c). 
From the data, we obtain the value of $D_r$ for the rotor as (1.4 $\rm{rad}^2/s$ ) by using 
a relation $MSD=2Dt+V^2t^2$ in one dimensions.

We report the self-propulsion of Janus particles under a wide laser irradiation. The mechanism of active motion is explained by the self-thermophoresis, which is	related to photophoresis \cite{arnold:1982} by photothermal effect. The direction of motion is determined by the polarity of a Janus particle in contrast to the phoretic motions guided by the external fields. The external field induced self-propulsion gives us a new way to power and control micro-machines. The basic principle is similar to the Crookes radiometer \cite{21}, but the scale is much more reduced.

We acknowledge the useful discussions with C.K. Chan and H. Wada. This works is supported by Grant-in-Aid for Scientific Research (18068005) from MEXT Japan.


\begin{thebibliography}{}
\bibitem{22}
Moderate distribution in Fig.3(e) is due to the fact that gold side is more often faced to the upper surface when reversed motion coincides with scattering force.
\end{thebibliography}

\end{document}